\documentclass[aps,prl,reprint,tightenlines,amssymb,amsmath,amsfonts,superscriptaddress,nofootinbib,twocolumn,showpacs,eqsecnum]{revtex4-1}

\usepackage{epsfig}
\usepackage{graphics}
\usepackage{graphicx}
\usepackage{amsmath,amssymb,mathrsfs}
\usepackage{amsfonts}
\usepackage[usenames,dvipsnames]{xcolor}
\usepackage{wasysym}
\usepackage{times}
\usepackage{mathptmx}
\usepackage{gensymb}
\usepackage{appendix}
\usepackage{listings}
\usepackage{listings}
\usepackage{url}
\usepackage[normalem]{ulem} 
\usepackage{alltt}
\usepackage[colorlinks]{hyperref}
\usepackage{longtable}
\usepackage{enumitem}
\usepackage{comment}
\setlist{nosep}

\definecolor{dodgerblue}{HTML}{1E90FF}
\definecolor{viennared}{HTML}{DA0A14}
\definecolor{ctorange}{HTML}{FF6C0C}
\definecolor{wales}{HTML}{ff0038}

\hypersetup{
     colorlinks=true,
     linkcolor=dodgerblue,
     filecolor=dodgerblue,
     citecolor = dodgerblue,      
     urlcolor=dodgerblue,
     }

\newcommand{\UoB}{School of Physics and Astronomy and Institute for Gravitational Wave Astronomy, University of Birmingham, Edgbaston, Birmingham, B15 9TT, United Kingdom}
\newcommand{\GRAPPA}{GRAPPA, Anton Pannekoek Institute for Astronomy and Institute of High-Energy Physics, University of Amsterdam, Science Park 904, 1098 XH Amsterdam, The Netherlands}
\newcommand{\DeltaITP}{Delta Institute for Theoretical Physics, Science Park 904, 1090 GL Amsterdam, The Netherlands}
\newcommand{\UIB}{Universitat de les Illes Balears, Crta. Valldemossa km 7.5, E-07122, Palma, Spain}

\begin{document}

\title{Gravitational-Wave Asteroseismology with Fundamental Modes from Compact Binary Inspirals}

\author{Geraint Pratten}
\email{gpratten@star.sr.bham.ac.uk}
\affiliation{\UoB}
\affiliation{\UIB}

\author{Patricia Schmidt}
\email{pschmidt@star.sr.bham.ac.uk}
\affiliation{\UoB}

\author{Tanja Hinderer}
\email{t.hinderer@uva.nl}
\affiliation{\GRAPPA}
\affiliation{\DeltaITP}

\begin{abstract}
The first detection of gravitational waves (GWs) from the binary neutron star (NS) inspiral GW170817 has opened a unique channel for probing the fundamental properties of matter at supra-nuclear densities inaccessible elsewhere in the Universe. 
This observation yielded the first constraints on the equation of state (EoS) of NS matter from the GW imprint of tidal interactions. Tidal signatures in the GW arise from the response of a matter object to the spacetime curvature sourced by its binary companion. They crucially depend on the EoS and are predominantly characterised by the  tidal deformability parameters $\Lambda_{\ell}$, where $\ell=2,3$ denotes the quadrupole and octupole respectively. As the binary evolves towards merger, additional {\textit{dynamical}} tidal effects become important when the orbital frequency approaches a resonance with the stars' internal oscillation modes. Among these modes, the fundamental ($f_\ell$-)modes have the strongest tidal coupling and can give rise to a cumulative imprint in the GW signal even if the resonance is not fully excited. Here we present the first direct constraints on fundamental oscillation mode frequencies for GW170817 using an inspiral GW phase model with an explicit dependence on the $f$-mode frequency and without assuming any relation between $f_\ell$ and $\Lambda_\ell$. We rule out anomalously small values of $f_\ell$ and, for the larger companion, determine a lower bound on the $f_2$-mode ($f_3$-mode) frequency of $\geq 1.39$ kHz ($\geq 1.86$ kHz) at the 90\% credible interval (CI).
We then show that networks of future GW detectors will be able to measure $f$-mode frequencies to within tens of Hz from the inspiral alone. Such precision astroseismology will enable novel tests of fundamental physics and the nature of compact binaries. 
\end{abstract}

\date{\today}

\begin{flushright}
LIGO-P1900131
\end{flushright}

\pacs{
04.80.Nn, 
95.85.Sz, 
97.80.-d   
04.30.Db, 
04.30.Tv,  
97.60.Jd, 
26.60.Kp  
}

\maketitle 

\section{Introduction}
Gravitational waves from inspiraling binary NSs carry characteristic information not only about the stars' masses and spins but also about the nature of the supra-dense matter in their interiors, whose properties have been a longstanding frontier of subatomic physics and astrophysics. The GW observation of the NS inspiral GW170817~\cite{GW170817} enabled the first constraints on NS matter via the EoS-dependent tidal deformability parameters $\Lambda_\ell$ for each object~\cite{Flanagan:2007ix, Wade:2014vqa, GWTC1}.
Tidal effects encapsulated in $\Lambda_\ell$ mainly describe the behavior in the early inspiral, when the stars respond adiabatically to their companion's tidal fields~\cite{Flanagan:2007ix}. However, as the GW-driven evolution proceeds, the variations in the tidal fields occurring on the orbital timescale can approach a resonance with the internal oscillation modes of the star giving rise to new characteristic signatures in the GWs associated with the specific modes. For the fundamental $f_\ell$-modes, such \emph{dynamical} effects lead to an $f$-mode frequency-dependent amplification of tidal imprints in the GW signal that starts to accumulate long before the resonance~\cite{Hinderer:2016eia,Steinhoff:2016rfi,Schmidt:2018aa}. In General Relativity (GR), and for a range of proposed EoSs for NSs, the $f_\ell$-mode frequencies are empirically found to be related to $\Lambda_\ell$ through approximate universal relations (URs)~\cite{Chan:2014kua}. However, not imposing such NS-based URs opens up the possibility of using GW observations to perform novel parameterized tests of GR, for understanding the fundamental properties of ultra-dense matter, and, especially, for testing for the existence of exotic compact objects such as boson stars or gravastars~\cite{Barack:2018yly}. 

\begin{figure*}[t!]
\includegraphics[width=\columnwidth]{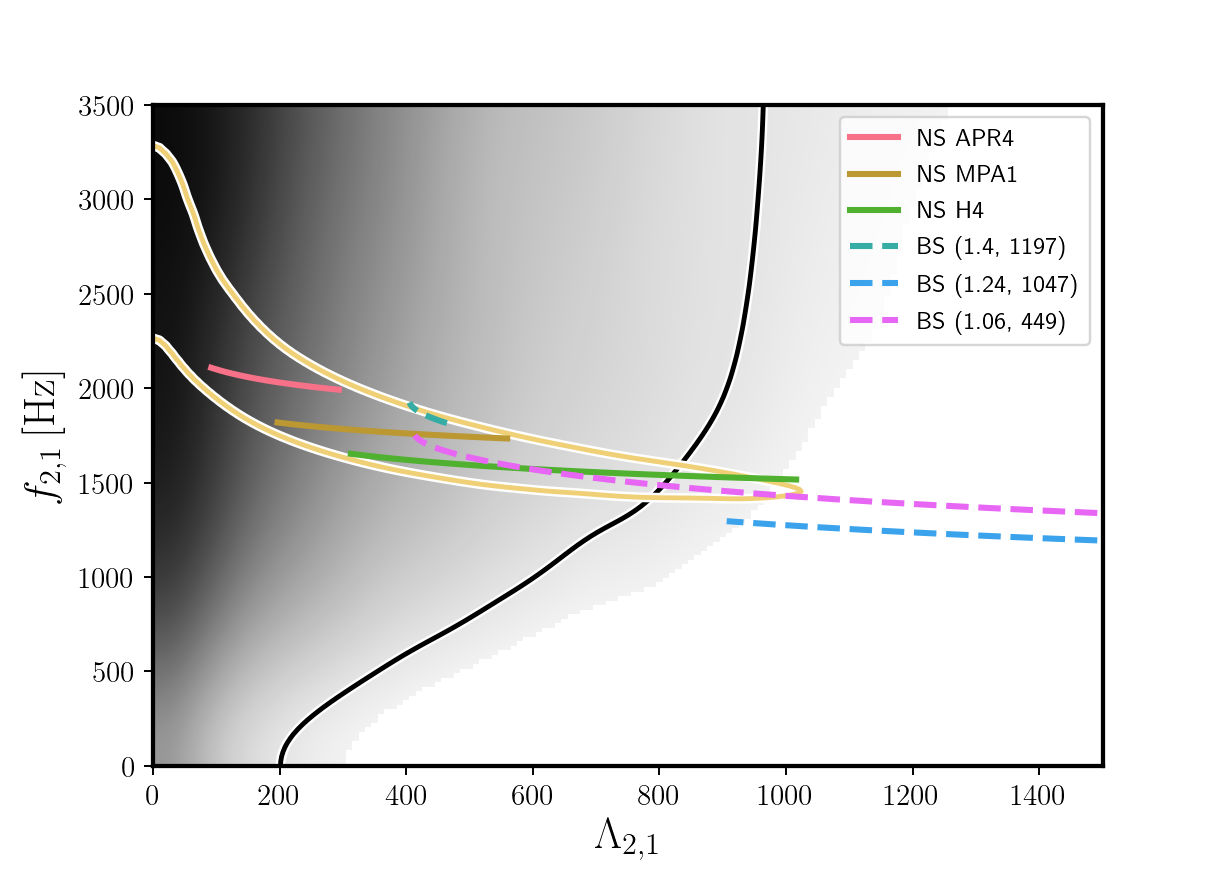}
\includegraphics[width=\columnwidth]{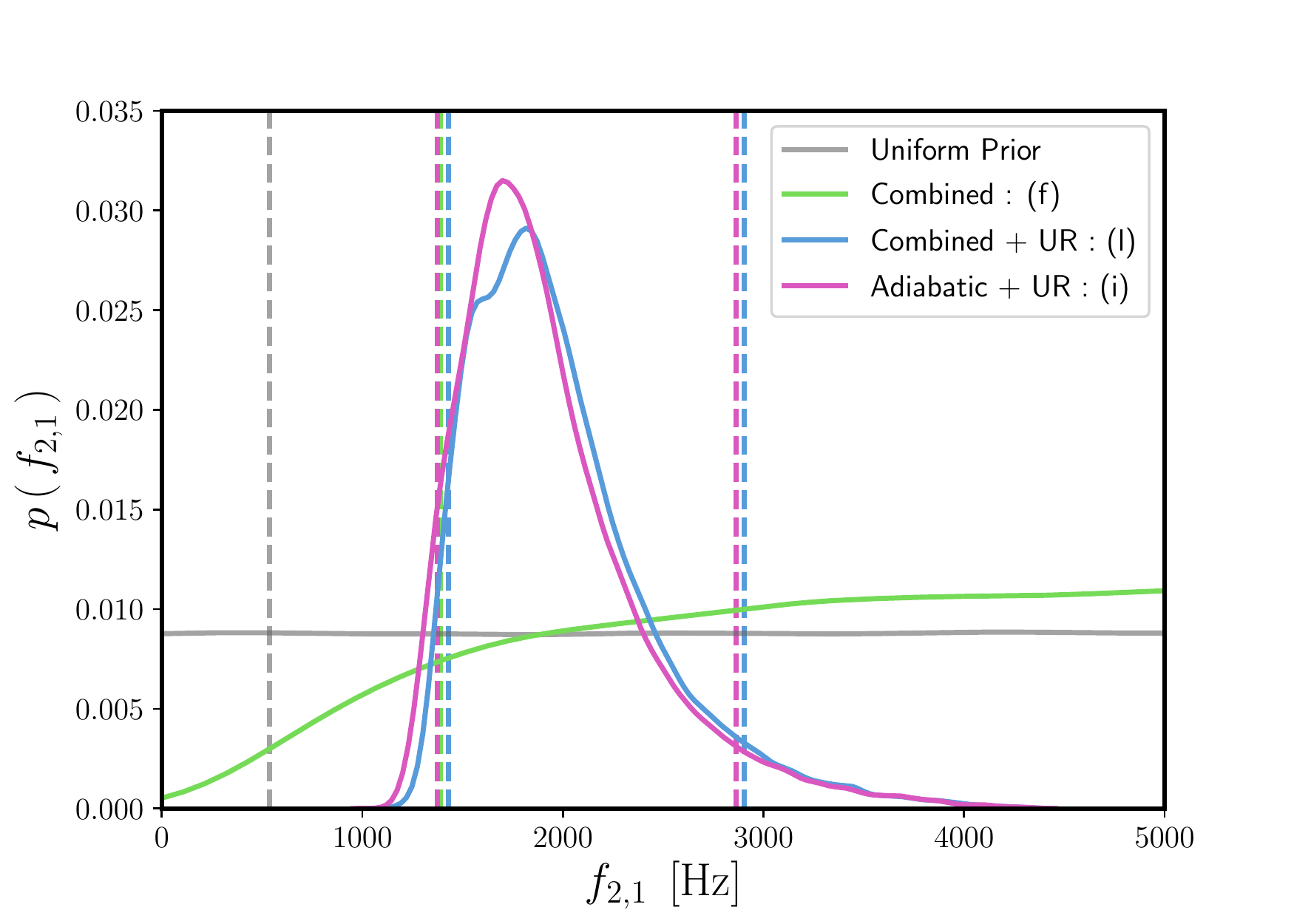}
\caption{\textbf{Results for the larger-mass object in GW170817.} Left: Two-dimensional PDF for the $f_2$-mode frequency and tidal deformability. The solid lines correspond to the $90\%$ credible regions. The posteriors are overlaid with UR predictions for three EoS for NSs (coloured solid curves), and three massive BSs (coloured dashed curves) denoted $(m_b/m_n, \lambda_b)$, with $m_n=1.675\times 10^{-27}$kg being the neutron mass, where all curves are restricted to the 90\% interval of the component mass posterior, $m_1 \in [1.37, 1.63] \,\, {\rm M}_\odot $. The black curve corresponds to the analysis in which $f_{2,A}$ is treated as an independent parameter and the yellow one to imposing the universal relations, i.e. fixing $f_{2,A}$ given $\Lambda_{2,A}$. Right: Marginalised 1D PDF for the $f_2$-mode frequency of the larger-mass companion in GW170817. We show results for the following three waveform models as listed in Tab.~\ref{tab:bounds}: (i) purely adiabatic tides with URs imposed (pink), (l) adiabatic and dynamical tides with UR imposed (blue) and (f) adiabatic and dynamical tides without UR assumed (green). The dashed lines indicate the corresponding 90\% lower bound (green) or CI (pink and blue).}
\label{fig:fL}
\end{figure*}

\section{Methodology}
\subsection{Parameter Estimation}
We perform coherent Bayesian parameter estimation on the publicly available GW170817 data \cite{GWOSC,Vallisneri:2014vxa}. We use the nested sampling algorithm implemented in \textsc{LALInference} \cite{skilling2006nested,Veitch:2014wba} to evaluate the posterior probability distribution
\begin{align}
    p(\vec{\lambda} | \vec{d}) &= \frac{ \mathcal{L}(d | \vec{\lambda}) \pi (\vec{\lambda}) }{\mathcal{Z} (\vec{d})} ,
\end{align}
\newline
where $\mathcal{L}(\vec{d} | \vec{\lambda} )$ is the likelihood of the data given the parameters $\vec{\lambda}$, $\pi (\vec{\lambda})$ the prior distribution for $\vec{\lambda}$ and $\mathcal{Z}$ the marginalized likelihood or evidence. Under the assumption of stationary Gaussian noise, the likelihood of obtaining a signal $h$ in the data $\vec{d}$ is given by
\begin{align}
   \mathcal{L}(\vec{d} | \vec{\lambda} ) &\propto \exp \left[ - \frac{1}{2} \displaystyle\sum_{k} \langle h_k^{M} (\vec{\lambda}) - d_k | h^M_k (\vec{\lambda}) - d_k \rangle \right] ,
\end{align}
\newline
where the $k$-th detector output is $d_k (t) = n_k (t) + h^M_k (t ; \vec{\lambda})$, $n_k (t)$ the noise and $h^M_k (t)$ the measured strain incorporating calibration uncertainty \cite{GWTC1}. For a single detector, the noise weighted inner product is given by
\begin{align}
    \left\langle a | b \right\rangle &= 4 \mathrm{Re} \int_{f_{\rm{low}}}^{f_{\rm{high}}} \frac{\tilde{a}(f) \tilde{b}^{\ast} (f)}{S_n (f)} \, df ,
\end{align}
\newline
where $\tilde{a}(f)$ and $\tilde{b}(f)$ denote the Fourier transform of the real-valued functions $a(t)$ and $b(t)$ respectively and $S_n (f)$ is the power spectral density (PSD) of the detector. In our analysis we adopt a a low-frequency cutoff of $f_{\rm{low}} = 23$ Hz and a high-frequency cut-off of $f_{\rm{high}} = 2048$ Hz. 

The choice of priors used in our analysis is templated on the choices made in Ref.~\cite{GWTC1}. For the dimensionless tidal deformabilities we use uniform priors with $\Lambda_{2,A}^{\rm prior} \in \left[ 0, 5000 \right]$ and $\Lambda_{3,A}^{\rm prior} \in \left[ 0 , 10^4 \right]$. Similarly, where appropriate, we adopt uniform priors on the dimensionless angular $f$-mode frequencies $\Omega_{\ell,A} = G m_A \omega_{\ell, A}/c^3$. We choose $\Omega_{2,A} \in \left[ 0, 0.5 \right]$ and $\Omega_{3,A} \in \left[ 0, 1.0 \right]$. These priors are decidedly conservative in order to remain as agnostic as possible. The upper limit imposed on $\Omega_{\ell,A}$ extends beyond the limit of $\sim 0.18$ implied by URs for $\Lambda_{\ell,A} \rightarrow 0$ \cite{Chan:2014kua}. Alternative prior choices can be considered but are beyond the scope of the analysis presented here.

In addition, there is some freedom associated to how we sample in the intrinsic parameters. Following \cite{Abbott:2018wiz}, we let the adiabatic tidal parameters $\Lambda_{2,A}$ vary independently and considered two different prescriptions for incorporating dynamical tides. In the first prescription, we let the $f$-mode frequencies vary independently, being treated as a free parameter to be constrained by the data. This allows us to place a lower bound on the $f$-modes avoiding any \textit{a priori} assumptions on the validity of the universal relations. In the second prescription, we can fix the $f$-mode frequencies by imposing universal relations, though we note that the dynamical tides terms still explicitly contribute to the log-likelihood in contrast to a purely adiabatic waveform. In addition, imposing the universal relations allows us to effectively reduce the dimensionality of the parameter space whilst still incorporating dynamical tidal effects. 

\subsection{Waveform Model}
In this study, we analysed GW170817 data using a fixed point particle baseline ($\Psi_{\rm pp}$) (see~\cite{Abbott:2018wiz} and Refs. therein for details) but incorporating adiabatic and dynamical tidal effects to varying post-Newtonian (PN) orders. The aim of this approach is to gauge the impact of systematics in the modelling of tidal effects on the estimation and bounds for the $f$-mode frequencies. 

The leading order adiabatic tidal effects first enter the phase at $5$PN order \cite{Flanagan:2007ix} and can be characterised by a single tidal deformability parameter $\tilde{\Lambda}$, which is a mass weighted average of the tidal deformabilities of the constituent compact objects $\Lambda_{2,A}$. The effects of adiabatic tidal deformations on the phase have been calculated to $6$PN in \cite{Vines:2011ud} and up to 7.5PN in \cite{Damour:2012yf}, where contributions from higher multipoles are omitted and a number of unknown terms appearing at $7$PN are neglected. In this paper, we also incorporate a recently derived closed-form expression for the dynamical tidal contribution to the phase \cite{Schmidt:2018aa}. This enables us to include the tidal excitation of the neutron stars fundamental oscillation modes at quadrupolar ($\ell = 2$) and octupolar ($\ell = 3$) order. In addition, we also include the octupolar adiababatic contribution to the phase as in \cite{Hinderer:2009ca}. The GW phase $\Psi$ computed in the PN approximation can therefore be written as the sum of the various contributions
\begin{align}
    \Psi (f) &= \Psi_{\rm{pp}} (f) + \Psi_{\rm{ad.}} (f) + \Psi_{\rm{dyn.}} (f) ,
\end{align}
\newline
where $\Psi_{\rm{pp}}$ is the point particle inspiral phase, $\Psi_{\rm{ad.}}$ the adiabatic tidal contributions to the phase and $\Psi_{\rm{dyn.}}$ the dynamical tidal contributions to the phase (``\texttt{fmtidal}'') from Ref.~\cite{Schmidt:2018aa}. We neglect quadrupole-monopole (QM) effects associated with the deformations of the NS under its own angular momentum \cite{Poisson:1997ha}, as for GW170817 the QM effects were found to be subdominant~\cite{Agathos:2015uaa}. However, the QM term can readily be incorporated into our phasing model. Future studies could further include a number of additional matter effects neglected here, and use more sophisticated waveform models beyond the PN-based approximations. 

Following \cite{Agathos:2015uaa}, we choose the waveform termination frequency as the minimum of ${\rm{min}} ( f_{\rm{ISCO}} , f_{\rm{contact}} )$, where $f_{\rm{ISCO}}$ is the innermost stable circular orbit and $f_{\rm{contact}}$ is a fiducial contact frequency.

\section{Constraints from GW170817}
In order to constrain $f$-mode frequencies in GW170817, we re-analyse the  publicly available strain data~\cite{GWOSC, Vallisneri:2014vxa} treating the tidal deformabilities $\Lambda_{\ell,A}$ \emph{and} the dimensionless angular $f$-mode frequencies $G m_A \omega_{\ell, A}/c^3$
of the $A$-th component object as independent parameters \emph{without} imposing the URs or any requirement that the two objects have the same EoSs. For comparison, we also repeat the analysis imposing the URs meaning that $f_\ell$ is no longer allowed to vary freely. For our analyses, we use the efficient post-Newtonian (PN) adiabatic TaylorF2 waveform model (see Refs. in~\cite{Abbott:2018wiz}) augmented with the $f$-mode dynamical tide phase developed in Ref.~\cite{Schmidt:2018aa}. We perform coherent Bayesian inference using the nested sampling algorithm implemented in \textsc{LALInference}~\cite{Veitch:2009hd, Veitch:2014wba}. We adopt the narrow spin priors of \cite{Abbott:2018wiz} and impose uniform priors on the $f$-mode frequencies.
We demonstrate that PN systematics do not dominate the lower bound by performing the analysis using different combinations of adiabatic and dynamical tide contributions, as summarised in Tab.~\ref{tab:bounds}. Note that the results quoted below will be taken from datasets (f) and (l) unless otherwise stated.

Figure~\ref{fig:fL} summarises our results for the larger-mass ($m_1$) object; we find similar results for the lower-mass ($m_2$) companion resulting in the same conclusions. 
The left panel of Fig.~\ref{fig:fL} shows the joint posterior probability distribution function (PDF) for the quadrupolar $f$-mode frequency $f_2=\omega_2/(2\pi)$ and corresponding tidal deformability $\Lambda_2$. The black solid curved indicates the 90\% credible region (CR) of our novel analysis with independent parameters, while the yellow curve is the 90\% CR obtained when imposing URs for NSs. 

The colored solid curves correspond to predictions from EoS models for NSs with increasing stiffness known as APR4~\cite{Akmal:1998cf}, MPA1~\cite{Muther:1987xaa}, and H4~\cite{Lackey:2005tk}; as an example of an exotic object we also show predictions for nonrotating boson stars\footnote{The $f$-mode frequencies computed in Ref.~\cite{Flores:2019iwp} use an approximate effective EoS and are used here solely for illustration as more complete results for cases relevant here are not available in the literature.}(dashed colored curves)~\cite{Sennett:2017etc,Flores:2019iwp}. Boson stars (BSs) are condensates of a complex scalar field $\Phi$ with a repulsive self-interaction described here by the potential $V=m_{\rm b}^2 |\Phi|^2+\frac{1}{2}\lambda_b|\Phi |^4$, where $m_b$ is the boson mass and $\lambda_b$ characterizes the strength of the self-interaction. 
We emphasise that the PDFs derived by imposing URs, as shown by the yellow curve in Fig.~\ref{fig:fL}, are only valid for binary neutron stars. In general, exotic compact objects and objects in modified theories of gravity may not obey the same set of URs, highlighting the utility of our approach. Figure~\ref{fig:fL} illustrates that measuring both the $f$-mode frequency and $\Lambda$ independently enables us to place additional constraints on the nature of the compact objects. 

Complementary to the joint posterior, the right panel of Fig.~\ref{fig:fL} shows the marginalised one-dimensional PDF of $f_{2}$ for the larger-mass companion.  
Without imposing the UR for NSs (green curve) on the dynamical tides terms, we can rule out anomalously small values of the $f$-mode frequency and place a meaningful lower bound on $f_2$ indicated by the green dashed vertical line. This can be understood from the scaling behaviour of the dynamical tides phase contribution which is proportional to $\Lambda_{\ell} f_{\ell}^{-2}$, implying that small values of $f_{\ell}$ result in hyper-excited dynamical tides that are inconsistent with the data. This is reflected in the shape of the posterior (green curve), where we see a dramatic drop in posterior support as $f \rightarrow 0$.
Conversely, increasing the fundamental frequency leads to a suppression of dynamical tidal effects and the waveform becomes indistinguishable from the adiabatic limit. The dynamical $f$-mode effects become most important at high frequencies where the detectors are less sensitive, thus the $f$-mode frequency could not be fully resolved and no upper bound can be determined as seen from the plateau at high frequencies. 
When URs are assumed, the posteriors on the $f$-mode frequencies yield an upper bound and are much narrower (blue curve), but most importantly the 90\% lower bound is consistent with our previous more general findings. 
Additionally, we also give the results with a purely adiabatic tidal phase model (pink curve), entirely omitting the frequency-dependent dynamical tidal enhancement from the GW model and reconstructing the posterior from the URs. We find that the lower bound is consistent with our estimate using the dynamical tides model and in broad agreement with \cite{Wen:2019ouw} -- this is not surprising as the strongest constraint stems from the tidal deformability measurement which would be inconsistent with very large dynamical tides (i.e., $f_2\rightarrow 0$). 

In Tab.~\ref{tab:bounds} we give the 90\% lower bounds on the quadrupolar $f$-mode frequency measured from GW170817 for various combinations of adiabatic and dynamical tidal phase contributions with and without assuming URs. We follow \cite{2017arXiv170309701H} and construct the bootstrapping estimate of the standard error on the lower limit, finding $f_{2,1}^{\rm low} = 1390 \pm 65 $ Hz for the larger mass. In particular, we note that the systematics between different the PN contributions given in Tab.~\ref{tab:bounds} are all within the bootstrapping errors estimated here, demonstrating robustness of the result.

In addition to the dominant quadrupole effects, we can also place weak constraints on octopolar tidal parameters. However, such higher multipole tidal interactions are subdominant and hence even more difficult to constrain from the data. For the larger component, the $90$\% credible interval on $\Lambda_{3,1}$ is $[410, 9404]$ Hz and the $90$\% lower bound on $f^{\rm low}_{3,1}$ is $1857 \pm 117$ Hz. Similarly, for the smaller component we find $\Lambda_{3,2} \in [466, 9446]$ Hz and $f^{\rm low}_{3,2} = 1619 \pm 98$ Hz. Note that the upper limits on $\Lambda_{3,A}$ are prior-dominated.

\begin{table}[t!]
    \centering
    \begin{tabular}{c|c|c}
    Tidal Phase Model & $f_{2,1}$ [kHz] & $f_{2,2}$ [kHz] \\
    \hline
    (a) 6.5PN ad. + $f_2$ dyn. & 1.47 & 1.57   \\
    (b) 7.5PN ad. + $f_2$ dyn. & 1.43 & 1.59  \\
    (c) combined & 1.45 & 1.58 \\
    \hline
    (d) 6.5PN ad. + $f_2$ + $f_3$ dyn. & 1.40 & 1.49 \\
    (e) 7.5PN ad. + $f_2$ + $f_3$ dyn. & 1.37 & 1.47 \\
    (f) combined & 1.39 & 1.48 
     \\
     \hline
    \hline
    (g) 6.5PN ad. + URs & 1.36 - 2.83 & 1.42 - 3.08 \\
   (h) 7.5PN ad. + URs & 1.37 - 2.90 & 1.43 - 3.16  \\
    (i) combined & 1.38 - 2.86 & 1.43 - 3.12 \\
    \hline
    (j) 6.5PN ad. + $f_2$ dyn. + URs & 1.42 - 2.88 & 1.48 - 3.17 \\
    (k) 7.5PN ad. + $f_2$ + $f_3$ dyn. + URs & 1.44 - 2.92 & 1.50 - 3.18 \\
    (l) combined & 1.43 - 2.90 & 1.48 - 3.18 
    \end{tabular}
    \caption{\textbf{Marginalised 90\% lower bound or 90\% CI (credible interval) of the quadrupolar $f$-mode frequency for both companions of GW170817.} We list results for all considered combinations of adiabatic and dynamical tide contributions in the waveform model. For results which assume URs, we also provide the upper bounds. Cases where $f_3$ is included also include the 7PN adiabatic octupolar effects \cite{Hinderer:2009ca}.}
    \label{tab:bounds}
\end{table}

While our analysis enables us to place a constraint on the $f$-modes for GW170817 from below and rule out very low values of $f_\ell$, little information is gained overall due to the decreased sensitivity at high frequencies for the current GW detector network. 
However, future detector networks will have the potential to place substantially tighter constraints on fundamental modes as we demonstrate next. 

\section{Future Prospects}
In order to illustrate the feasibility of measuring $f$-modes from inspiral signals with future observatories, we consider a GW170817-like binary with component masses $m_A =\{1.475,1.26\} \,\, {\rm M}_\odot$, tidal deformabilities $\Lambda_{2,A}=\{183, 488\}$ and $f_2$-mode frequencies $f_{2,A}=\{2.04,1.94\}$ kHz based on the APR4 EoS~\cite{Akmal:1998cf}. 
At high signal-to-noise ratios (SNRs), we can use the linear signal approximation $h(\vec{\lambda}) = h(\vec{\lambda}_0) + \partial_i h \Delta \lambda^i + \cdots$, where $\Delta \lambda^i = \vec{\lambda} - \vec{\lambda}_0$, to estimate the multivariate PDF about the true binary parameters $\vec{\lambda}_0$ as $p(\vec{\lambda}_0) \sim e^{-(1-\mathcal{M})\rho^2}$ \cite{Cutler:1994ys,Chatziioannou:2017tdw}. Here $\rho$ is the SNR and $1-\mathcal{M}$ is the mismatch between two waveforms. For $h$ we use the TaylorF2 approximant with 6.5PN adiabatic tidal effects and only include the $f_2$ dynamical tides contribution; we do not impose URs. 
\begin{figure}[t!]
\includegraphics[width=\columnwidth]{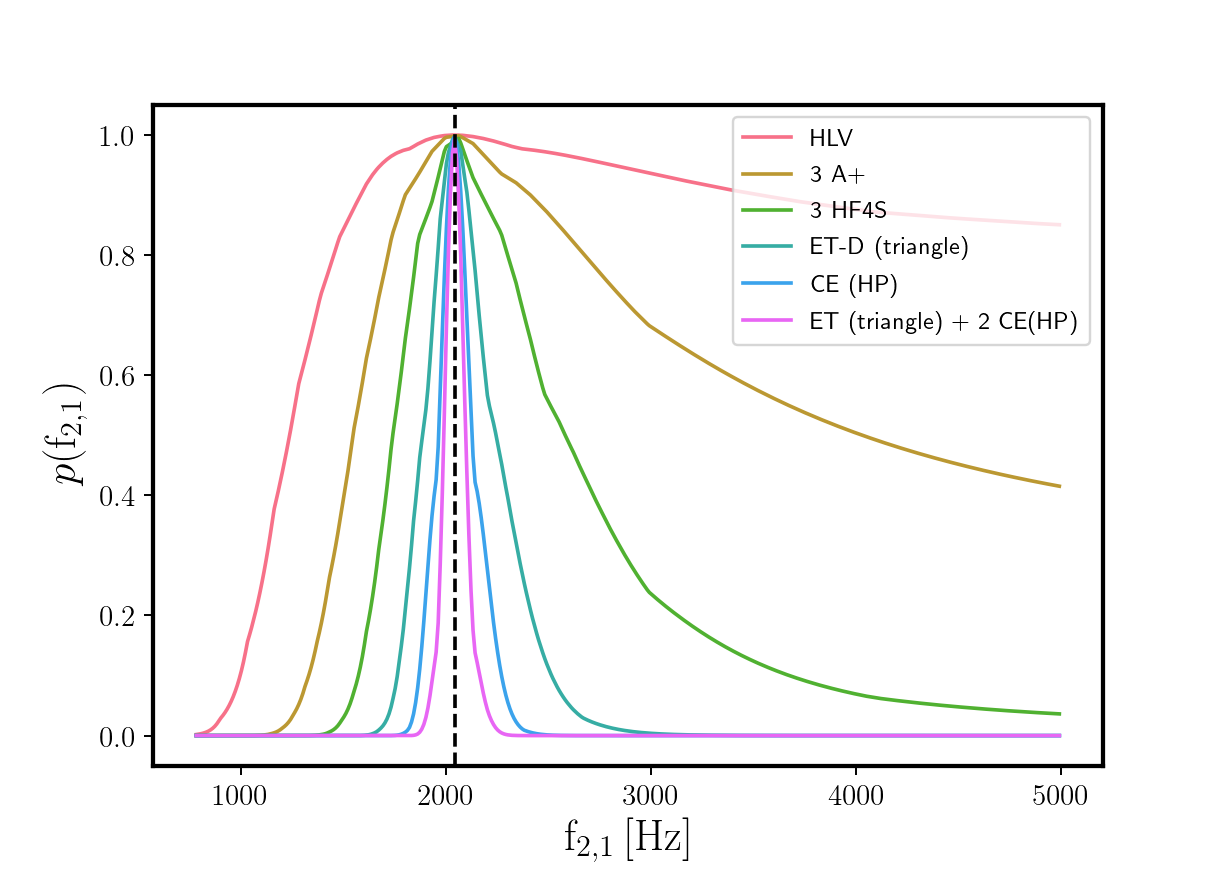}
\caption{\textbf{Approximate 1D posterior probability for the $f_2$-mode frequency of a GW170817-like binary at 40 Mpc in different detector networks.} The vertical dashed line indicates the true value of $f_{2}=2.04$ kHz for the larger mass. The detector networks considered are: LIGO-Virgo at design sensitivity (HLV), three A+ detectors~\cite{Miller:2014kma, APlusDesign}, three 4km L-shaped LIGO detectors with improved high-frequency sensitivity~\cite{Miao:2017qot,Martynov:2019gvu} (HF4S), one triangular Einstein Telescope D-configuration, one Cosmic Explorer (CE) and a network consisting of two CEs and one ET-D.}
\label{fig:posterior}
\end{figure}
Figure~\ref{fig:posterior} shows the approximate one-dimensional PDF for the $f_{2}$-mode frequency of the larger-mass object for an optimally-oriented GW170817-like binary at 40 Mpc in different detector networks. The result for the LIGO-Virgo detector network at design sensitivity~\cite{TheLIGOScientific:2014jea, TheVirgo:2014hva} shows little improvement over our actual measurement for GW170817 (compare to Fig.~\ref{fig:fL}), only allowing us to determine a lower bound. However, various future network configurations all enable us to place much tighter constraints on the $f$-mode frequencies, highlighting the potential of such detectors:
A network of three A+ detectors~\cite{Miller:2014kma, APlusDesign} will offer significant improvements over the Advanced LIGO-Virgo (HLV) network. In particular, we find that we can begin to distinguish between an adiabatic waveform and dynamical tides sourced by large $f$-mode frequencies. By optimizing current detectors at high frequencies~\cite{Miao:2017qot,Martynov:2019gvu} (HF4S), we find that we can start to place meaningful $90$\% lower \emph{and} upper limits on $f_2$ (green curve). 
A future 3G network consisting of one Einstein Telescope (ET) detector~\cite{Punturo:2010zz, Hild:2010id} and two Cosmic Explorer (CE) observatories~\cite{Evans:2016mbw} will allow for $1\sigma$-errors of only a few tens of Hz, making precision GW asteroseismology with inspiral signals possible.
This simplified calculation neglects correlations between intrinsic parameters and is limited by systematics of the waveform model; it should be viewed as a proof of principle that we can make meaningful measurements of fundamental mode oscillations from compact binary inspirals.
A detailed study on the measurability of $f$-modes in future GW detector networks will be presented in forthcoming work. 

\section{Conclusions}
In conclusion, we have presented the first direct constraints on the fundamental oscillation modes in GW170817 using a waveform model with dynamical tides and without assuming URs. We have demonstrated that meaningful measurements of $f$-mode dynamical tides are a unique possibility in a future GW detector network, opening the prospects for deriving fundamentally new science from the clean inspiral epoch that is complementary to the information from post-merger signals \cite{Bauswein:2011tp}.



\bibliography{References}

\section{Acknowledgments}
The authors thank Alberto Vecchio and Ben Farr for useful discussions and comments on the manuscript, and Denis Martynov and Haixing Miao for providing us with the sensitivity curve used in Fig. 2.
G. Pratten acknowledges support from the Spanish Ministry of Culture and Sport grant FPU15/03344, the Spanish Ministry of Economy and Competitiveness grants FPA2016-76821-P, the Agencia estatal de Investigaci\'on, the RED CONSOLIDER CPAN  FPA2017-90687-REDC, RED CONSOLIDER MULTIDARK: Multimessenger Approach for Dark Matter Detection, FPA2017-90566-REDC, Red nacional de astropart\'iculas (RENATA), FPA2015-68783-REDT, European Union FEDER funds, Vicepresid\`encia i Conselleria d'Innovaci\'o, Recerca i Turisme, Conselleria d'Educaci\'o, i Universitats del Govern de les Illes Balears i Fons Social Europeu, Gravitational waves, black holes and fundamental physics.
P. Schmidt acknowledges support from the Netherlands Organisation for Scientific Research (NWO) Veni grant no. 680-47-460. T. Hinderer acknowledges support from the DeltaITP and NWO Projectruimte grant GW-EM NS. This research has made use of data, software and/or web tools obtained from the Gravitational Wave Open Science Center~\cite{GWOSC}, a service of LIGO Laboratory, the LIGO Scientific Collaboration and the Virgo Collaboration. LIGO is funded by the U.S. National Science Foundation. Virgo is funded by the French Centre National de Recherche Scientifique (CNRS), the Italian Istituto Nazionale della Fisica Nucleare (INFN) and the Dutch Nikhef, with contributions by Polish and Hungarian institutes. The authors are grateful for computational resources provided by the LIGO Laboratory -- Caltech Computing Cluster and supported by the National Science Foundation.

\section{Supplementary Information}

We provide the results for the smaller-mass object of GW170817 as well as additional information. In particular, we provide results summary plots for the octupolar tidal deformabilities and $f$-mode frequencies. 

\begin{figure*}
    \includegraphics[width=\columnwidth]{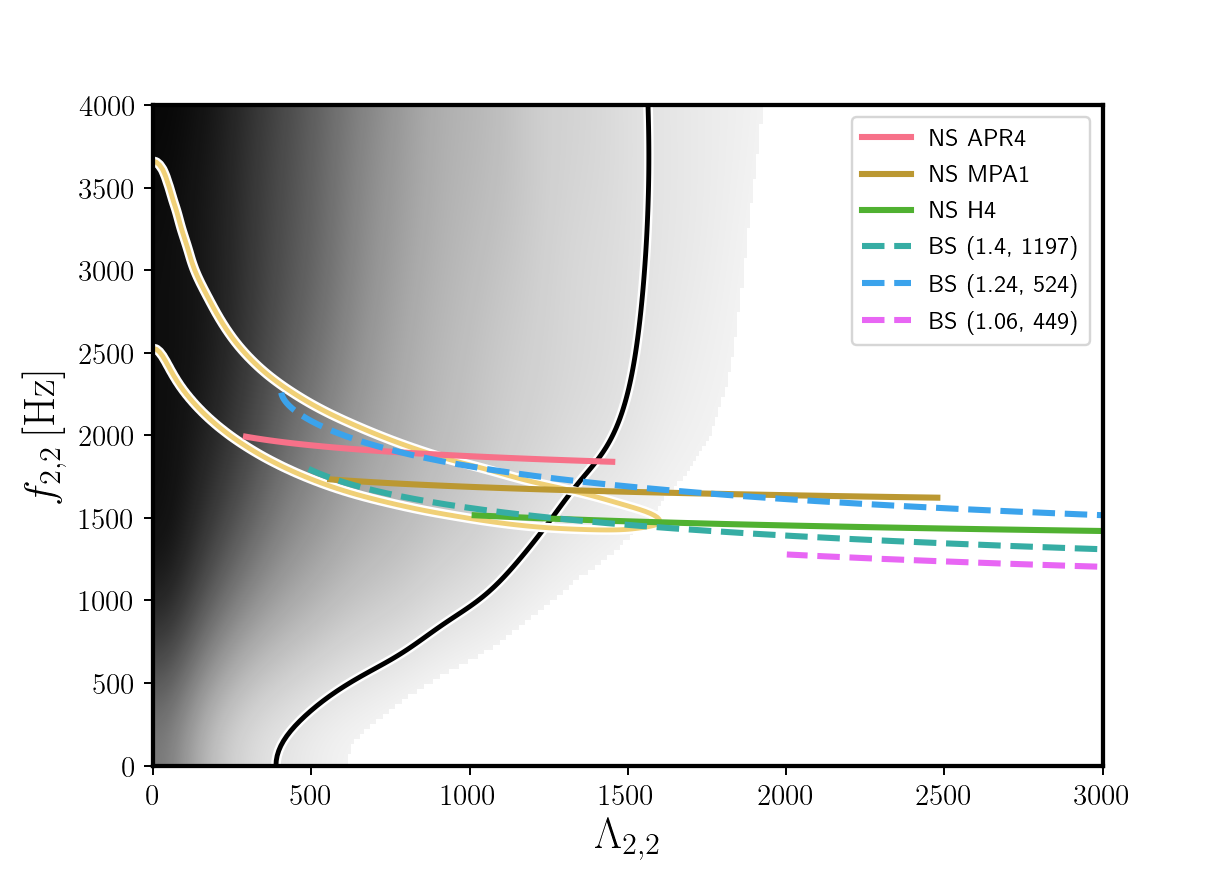}
    \includegraphics[width=\columnwidth]{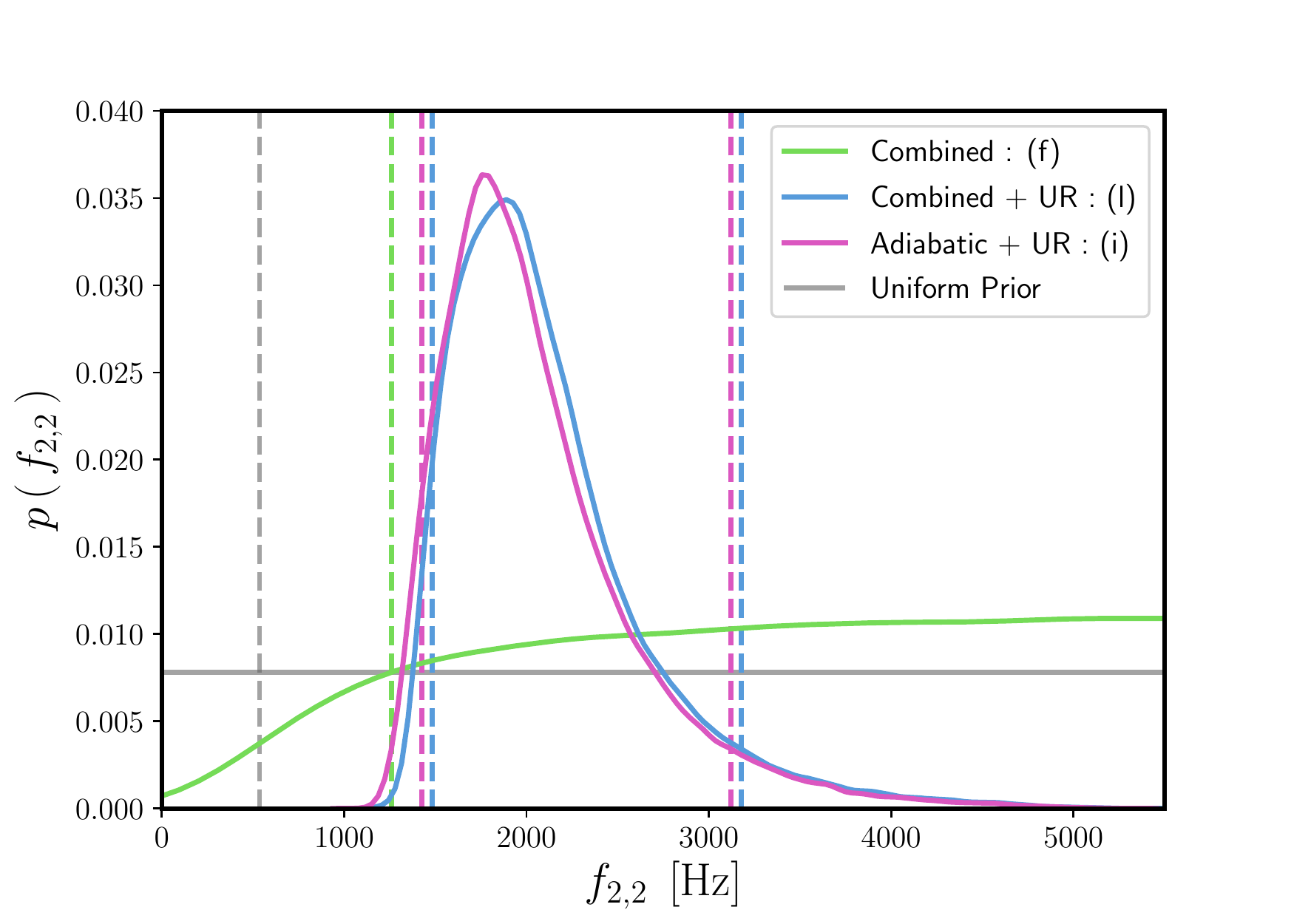}
    \caption{Complementary to Fig.~\ref{fig:fL} showing the results for the smaller-mass object $m_2$ of GW170817. Left: Two-dimensional PDF for the $f_2$-mode frequency and tidal deformability. The solid lines correspond to the $90\%$ credible regions. The posteriors are overlaid with UR predictions for three EoS for NSs (coloured solid curves), and three massive BSs (coloured dashed curves) denoted $(m_b/m_n, \lambda_b)$, with $m_n=1.675\times 10^{-27}$kg being the neutron mass, where all curves are restricted to the 90\% interval of the component mass posterior, $m_2 \in [1.04, 1.37] \,\, {\rm M}_\odot $. The black curve corresponds to the analysis in which $f_{2,A}$ is treated as an independent parameter and the yellow one to imposing the universal relations, i.e. fixing $f_{2,A}$ given $\Lambda_{2,A}$. Right: Marginalised 1D PDF for the $f_2$-mode frequency of the smaller-mass companion in GW170817. We show results for the following three waveform models as listed in Tab.~\ref{tab:bounds}: (i) purely adiabatic tides with URs imposed (pink), (l) adiabatic and dynamical tides with UR imposed (blue) and (f) adiabatic and dynamical tides without UR assumed (green). The dashed lines indicate the corresponding 90\% lower bound (green) or CI (pink and blue).}
    \label{fig:}
\end{figure*}

\begin{figure*}
    \centering
    \includegraphics[width=\columnwidth]{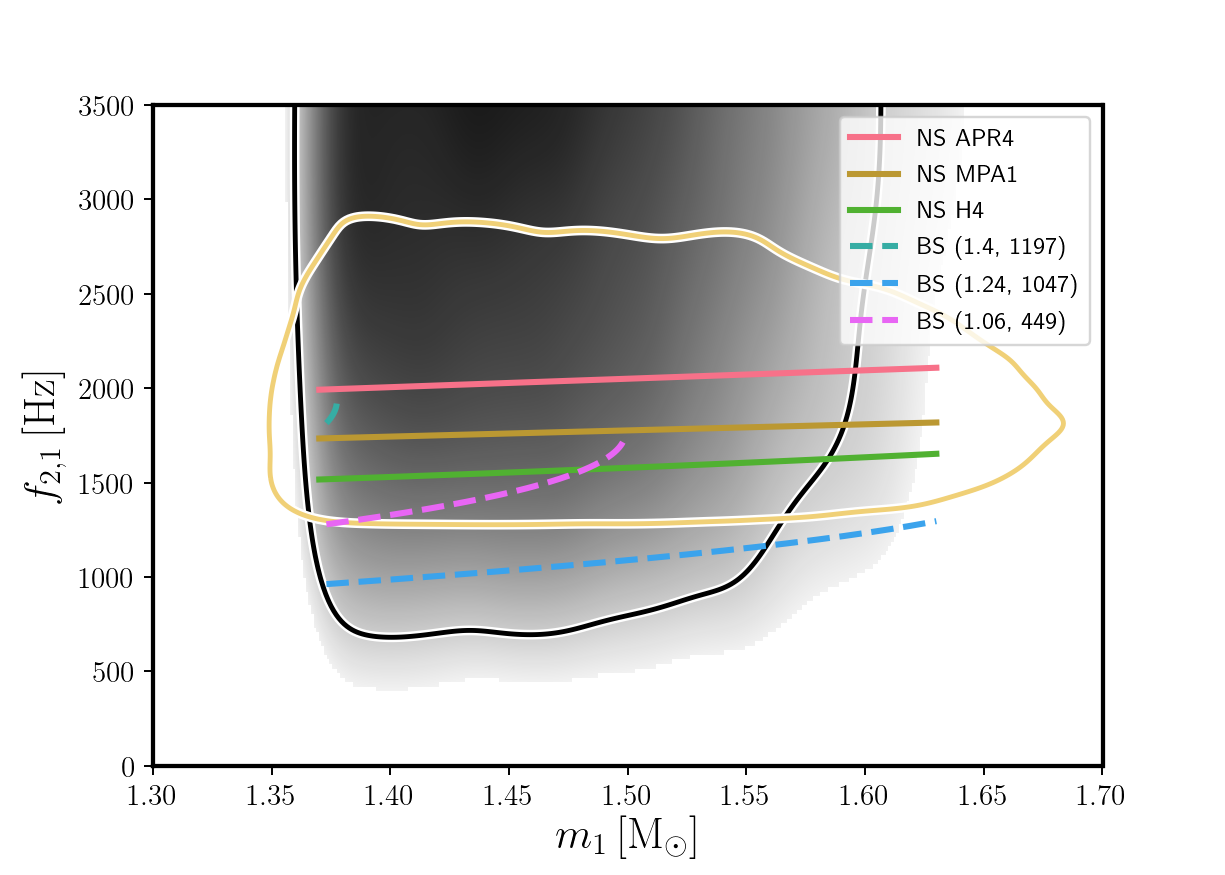}
    \includegraphics[width=\columnwidth]{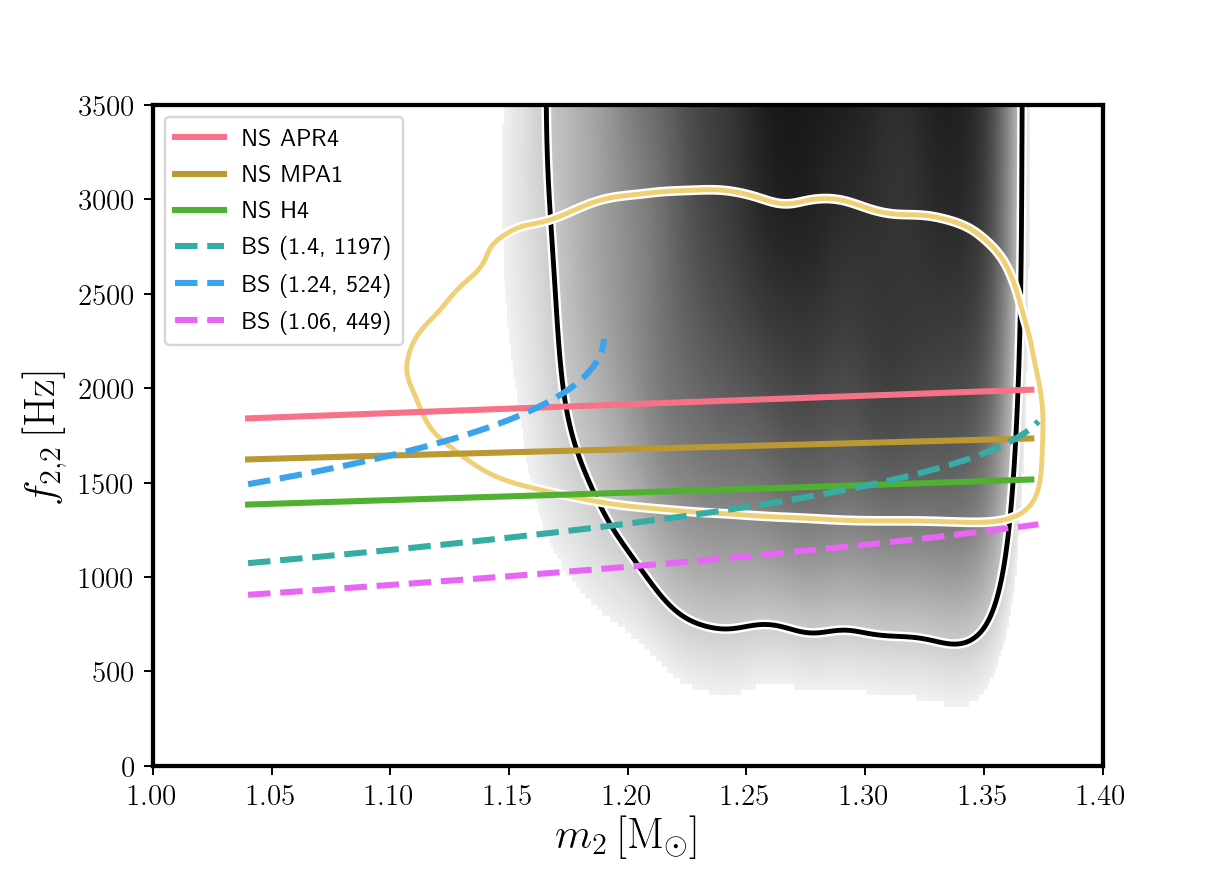}
    \caption{Joint two-dimensional PDF for component mass and quadrupolar $f$-mode frequency for GW170817. The results for the larger and smaller mass objects are shown in the left and right panel respectively. The information displayed here complements Fig.~\ref{fig:fL} in the main text by showing explicitly the information on the masses resulting from our analysis. Providing $f$-mode and mass information without reference to the tidal deformability is further useful for direct constraints as most calculations of compact-object oscillation modes are done within a quasi-normal modes framework that only gives the frequency but not $\Lambda_\ell$.}
    \label{fig:}
\end{figure*}

\begin{figure*}
    \includegraphics[width=\textwidth]{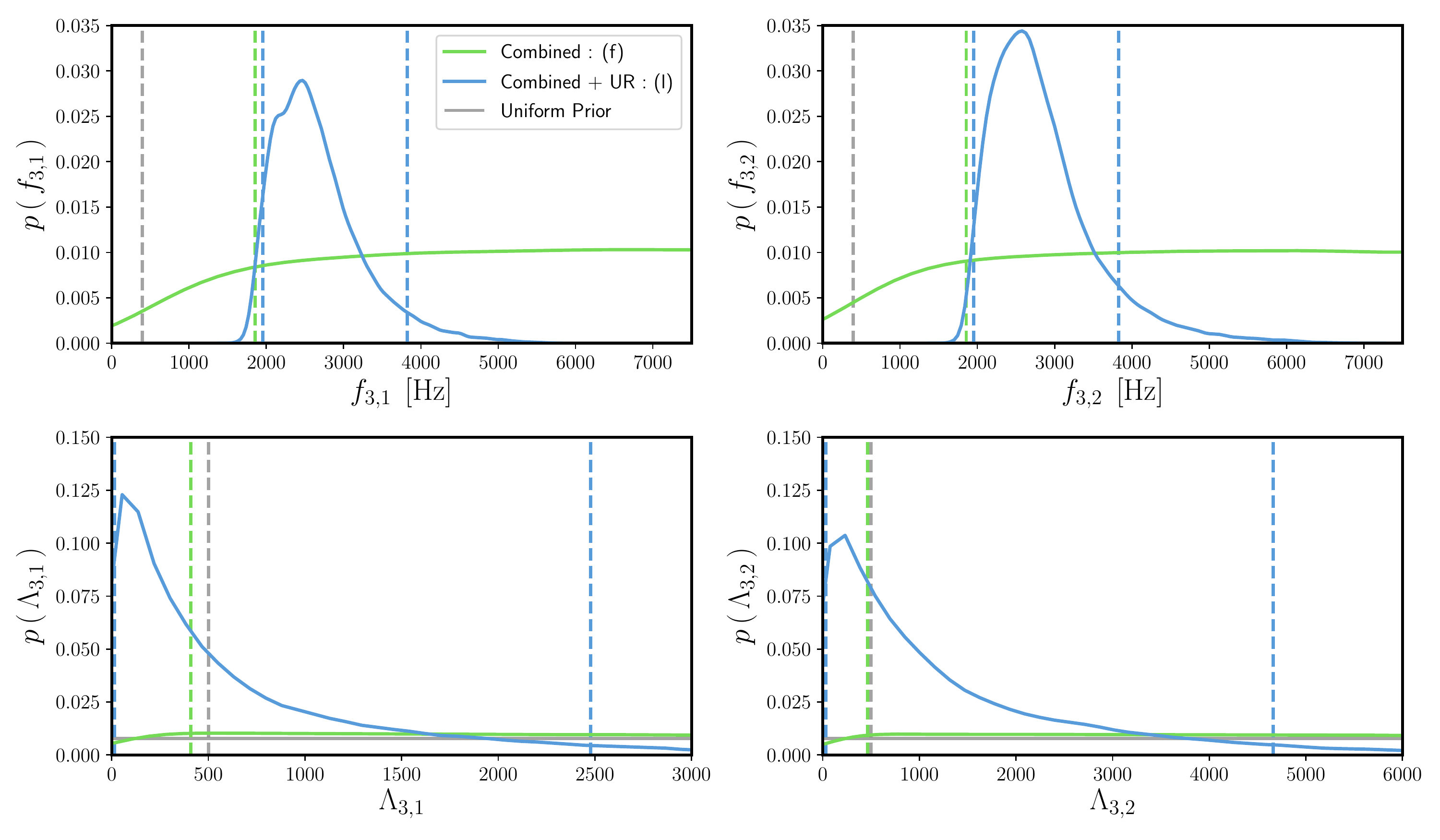}
    \caption{Constraints on the octupolar $f$-mode frequency ($f_3 = \omega_3/(2\pi)$) and octupoolar tidal deformability $\Lambda_{3,A}$ for both companions in GW170817. Note that for the unconstrained tidal deformabilities, our results are manifestly dominated by the prior with the posteriors demonstrating a slight shift towards smaller values of $\Lambda_{3}$, as predicted by the URs. As with the quadrupolar $f$-modes, the lower bound on $f_3$ is in agreement with the lower limit implied by imposing URs.}
    \label{fig:f3L3}
\end{figure*}

\begin{table}[h!]
    \centering
    \begin{tabular}{c|c}
    Network & SNR \\
    \hline
    HLV & 71 \\
    3 A+ & 145 \\
    3 HF4S & 170 \\
    ET-D & 788 \\
    CE & 1107 \\
    ET + 2 CE & 1753 
    \end{tabular}
    \caption{Network SNR of a GW170817-like source at 40 Mpc for all network configurations considered in Fig.~\ref{fig:posterior}.}
    \label{tab:SNRs}
\end{table}


\end{document}